\documentclass[traditabstract,letter]{aa}
\usepackage{txfonts}
\usepackage{natbib}
\bibpunct{(}{)}{;}{a}{}{,}
\usepackage[pdftex]{graphicx}
\usepackage{indentfirst}
\usepackage{amssymb}
\begin{document}
\title{Radial transport of refractory inclusions and their preservation in the dead zone}


\author{Emmanuel Jacquet\inst{1} \and S\'{e}bastien Fromang\inst{2,3} \and Matthieu Gounelle\inst{1}}

\institute{Laboratoire de Min\'{e}ralogie et de Cosmochimie du Mus\'{e}um (LMCM), CNRS \& Mus\'{e}um National d'Histoire Naturelle, UMR 7202, 57 rue Cuvier, 75005 Paris, France. \email{ejacquet@mnhn.fr} \and CEA, Irfu, SAp, Centre de Saclay, F-91191 Gif-sur-Yvette, France \and UMR AIM, CEA-CNRS-Univ. Paris VII, Centre de Saclay, F-91191 Gif-sur-Yvette, France}


\abstract{Calcium-aluminum-rich inclusions (CAIs) are the oldest solar system solids known in primitive meteorites (chondrites). They predate the other components by 1-2 Myr, and likely condensed within a short time interval, close to the Sun in the gaseous protoplanetary disk. Their preservation must counterbalance both the sunward drift caused by gas drag and the general inward motion of the entraining gas. We propose that an efficient outward transport of CAIs can be achieved by advection as a result of the viscous expansion of the disk, provided it is initially less than 10 AU in size, which we argue is plausible from both observational and theoretical points of view.  Gas drag would stop this outward motion within $10^5$ yr. However, by that time, a magnetically dead zone would have developed as gravitational instabilities fade away, which would trap CAIs for a significant fraction of the disk lifetime because of the reduced advection velocities. The dead zone would also prevent outward diffusion of subsequently condensed CAIs, contributing to their observed narrow age range. This preservation mechanism is independent of the outward transport scenario (before the dead zone formation) and a natural consequence of considering the source of turbulence in accretion disks.}

\titlerunning{Radial transport of CAIs}
\authorrunning{Jacquet et al.}

\maketitle

\section{Introduction}
 \textit{Chondrites} are the most primitive meteorites and provide an invaluable record of early solar system processes.They consist of 
millimeter-sized solids set in a fine-grained matrix, 
the most abundant of which are \textit{chondrules}, silicate spherules produced by rapid melting and cooling of precursor material
\citep[e.g.][]{ConnollyDesch2004}. Another noteworthy, albeit rarer component of chondrites are \textit{refractory inclusions},in particular submillimeter- to centimeter-sized \textit{calcium-aluminum-rich inclusions} (CAIs). Thermodynamic calculations suggest that they condensed at temperatures of $\sim$1400-1800 K from a gas of solar composition, although many have since melted  and/or partially evaporated.With an age of $\sim$4568 Myr \citep[e.g.][]{Bouvieretal2007}, CAIs are the oldest known solar system solids. Intriguingly, primary formation of CAIs---at least those of CV chondrites---seems restricted to an interval of a few $10^4$ yr \citep{Bizzarroetal2004}, while chondrule formation apparently began 1-2 Myr later and lasted $\sim$2 Myr \citep{Villeneuveetal2009}. Despite some variation in CAI mineralogy among chondrite groups, similarities indicate that all CAIs formed in a common environment, likely close to the Sun, and were then transported to chondrite-forming regions \citep[e.g.][]{McPherson2005}. Outward transport is also needed for two CAIs identified in dust returned from comet Wild 2 \citep[e.g.][]{Zolenskyetal2006}.

  Transport and/or production mechanisms of CAIs must have been efficient because their mass fraction in some chondrites---a few percent, e.g. \citet{Hezeletal2008}---is comparable with what \textit{in situ} condensation from a solar gas would have achieved. Moreover, these mechanisms must offset the sunward drift induced by gas drag on solids \citep{Weidenschilling1977} and the general inward motion of accreting gas, which partially entrains the solids. These combined phenomena typically entail a drift into the Sun within a few $10^5$ yr for millimeter-sized bodies. 

   Proposed CAI transport scenarii either involve jets such as the X-Wind model \citep{Shuetal2001}, or operate solely inside the disk, e.g. with turbulent diffusion \citep[e.g.][]{BockeleeMorvanetal2002,Cuzzietal2003,Ciesla2010}, gravitational instabilities \citep[GI;][]{Boss2004}, or meridional circulation---an hypothetical outward flow around the midplane \citep[e.g.][]{TakeuchiLin2002,Ciesla2009}. As originally formulated in the literature, jet models and meridional circulation should operate over most of the disk's lifetime and thus would not reproduce \textit{per se} the narrow age range of CV CAIs. Conversely, turbulent diffusion and GI would be efficient only in the early phases. As \citet{Cuzzietal2003} emphasize, disk evolution definitely matters.

  In this letter, we sketch a scenario (depicted in Fig. 1) of CAI transport and preservation in the disk: We adopt in Sect. 2 a small and massive initial disk (stage 'A' in Fig. 1). As described in Sect. 3, viscous heating enables CAI condensation sufficiently far from the Sun to have CAIs entrained outward by disk expansion (stage 'B'), before they are stopped by gas drag (stage 'C'). As GI fade away, a dead zone forms, where the magnetorotational instability (MRI) does not operate, in which CAIs are effectively trapped for a few Myr until chondrite accretion (stage 'D') because of reduced advection velocity. This is the subject of Sect. 4. The model and its limitations are discussed in Sect. 5.
\section{Disk model}  
\begin{figure}
\resizebox{\hsize}{!}{\includegraphics{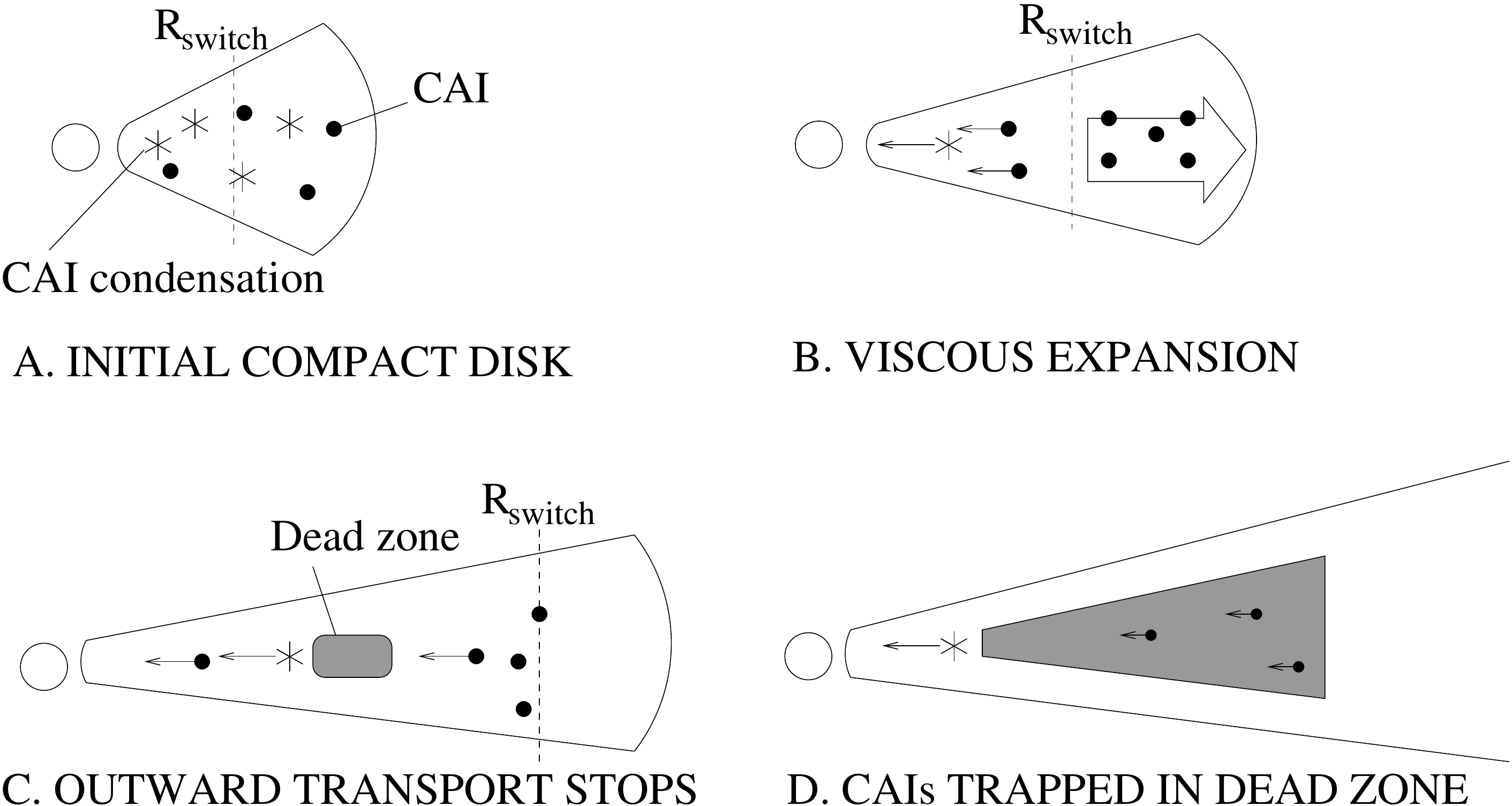}}
\caption{Cartoon of the proposed scenario: CAIs form in an initially compact and massive disk (A), which undergoes viscous expansion, entraining CAIs outward (B), building up a leaky reservoir that is eventually released at large heliocentric distances (C), while a dead zone emerges and traps CAIs until chondrite accretion (D).}
\end{figure}

Observations of class 0 objects have not yet unambiguously identified disks, suggesting that many disks may have begun by being very compact \citep[][and references therein]{Mauryetal2010,HennebelleCiardi2009}. The lower range of plausible values of the precollapse cloud's rotation velocity and size actually allow a centrifugal radius---the maximum stellocentric distance where infalling material lands on the equatorial plane---less than $\sim$10 AU \citep{Pickettetal1997,Zhuetal2010b}. Hydrodynamical simulations of protostellar collapse by \citet{Tscharnuteretal2009} result within $\sim 10^4$ yr in the buildup of a compact disk with a hot inner region extending to 4 AU. Moreover, magnetic braking during the collapse has been shown to be able to remove angular momentum efficiently enough to allow small disk sizes, depending on the relative inclination between angular momentum and the magnetic field threading the cloud \citep{HennebelleFromang2008,HennebelleCiardi2009}. 

  We therefore consider an early epoch where the disk size is a few AU, and adopt, for greater definition, the surface density profile of the self-similar model taken from \citet{LyndenBellPringle1974}, similarly to \citet{Cuzzietal2003}:
\begin{equation}
\Sigma (t,R) = \frac{M_D(t)}{2\pi R R_D(t)}e^{-\frac{R}{R_D(t)}},
\label{Sigmaprofile}
\end{equation}
with $M_D(t)$ and $R_D(t)$ the disk mass and characteristic radius, respectively, at time $t$. This model assumes that the viscosity $\nu=\alpha c_s^2/\Omega$ (with $c_s$ the isothermal sound speed, $\Omega$ the Keplerian angular velocity and $\alpha$ a dimensionless parameter) is \textit{time-independent} and proportional to the heliocentric distance $R$, and therefore may be written as 
\begin{equation}
\nu=5\times 10^{10}\:\mathrm{m^2/s}\:f_{\nu}\rm R_{\rm AU},
\label{nuprofile}
\end{equation}
 with $f_{\nu}$ a dimensionless factor, and $R_{\mathrm{AU}}$ the heliocentric distances expressed in AU. In the disk outer regions dominated by the irradiation temperature (here $\propto R^{-1/2}$, with nominal $T_{\mathrm{irr}}(1\:\mathrm{AU}) = 280\:\mathrm{K}$), our nominal $f_{\nu}=1$ corresponds to a constant $\alpha\!=\!10^{-2}$. $R_D$ increases linearly with time and satisfies
\begin{equation}
 t=\frac{R_D(t)^2}{3\nu(R_D(t))},
\label{time-radius}
\end{equation}
 the origin of times being a mathematical singularity where all mass resides at the center. Angular momentum conservation implies that $M_DR_D^{1/2}$ is constant. 
 We take a nominal $M_DR_D^{1/2}=0.5\:\rm M_{\odot}.AU^{1/2}$, a reasonable value because, in evolved phases, the disk, whose radius is $>10^2$ AU, may be expected to have a mass larger than that of the ``minimum mass solar nebula'' (MMSN), 0.01-0.02 $\rm M_{\odot}$ \citep{Hayashi1981}. Moreover, one-zone models of embedded accretion disks by \citet{Kratteretal2008} yield a disk-to-star mass ratio of a few tenths during the infall phase. 

  It is important to note, based on Eq. (\ref{time-radius}), that within $\sim$1 Myr the disk, though initially small, \textit{will} attain $10^2$-$10^3$ AU as observed for class II objects.
\section{Outward transport via viscous expansion} 
In the self-similar model, the vertically averaged radial
 gas velocity is \citep{LyndenBellPringle1974}
\begin{equation}
v_\mathrm{g}=-\frac{3}{\Sigma \sqrt{R}}\frac{\partial}{\partial R}\left(\sqrt{R}\Sigma\nu\right)=3\nu\left(\frac{1}{R_D}-\frac{1}{2R}\right),
\label{vg}
\end{equation}
where the second equality uses the surface density profile given by Eq. (\ref{Sigmaprofile}).
 Beyond the turnover radius  $R_{\mathrm{switch}}=R_D/2$
, $v_g$ is outward (positive). 

  In the inner disk, viscous dissipation generates a contribution to the local temperature that dominates that of irradiation by the Sun. Using \citet{Chambers2009}'s formulae, but sticking to the prescribed viscosity profile given by Eq. (\ref{nuprofile}), the former is
\begin{equation}
T_{\mathrm{vis}}=10^3\:\mathrm{K}\:\left(f_{\nu}\frac{\kappa}{0.5\:\mathrm{m^2/kg}}\right)^{1/4}\left(\frac{\Sigma}{10^4\:\mathrm{kg/m^2}}\right)^{1/2}R_{\rm AU}^{-1/2},
\end{equation}
 with $\kappa$ the (constant) opacity. 
 Initially, as shown in Fig. 2, the ``CAI factory'' \citep{Cuzzietal2003} extends beyond $R_{\mathrm{switch}}$, allowing CAIs to be advected outward from where they condensed (stage 'B' in Fig. 1) until $R_{\rm switch}$ exits the CAI factory on a timescale of $10^4-10^5$ yr (see Fig. 2). This alone would account for the narrow age range of CV CAIs.  

  The CAIs cannot be entrained arbitrarily far, however. The gas drag in the Epstein regime indeed induces a drift with respect to the gas \citep[e.g.][]{Weidenschilling1977}:
\begin{equation}
v_{\mathrm{drag}}=\sqrt{\frac{\pi}{8}}\frac{\partial P}{\partial R}\frac{\rho_sa}{\rho^2c_s},
\label{vdrag}
\end{equation}
with $P$, $\rho$ the gas pressure and density, and $\rho_s$, $a$ the CAI internal density and radius, respectively. $v_{\rm drag}$ is inward and roughly \textit{inversely} proportional to $\rho$. Thus, for a given CAI size, at sufficiently large heliocentric distances, $v_{\rm drag}$ dominates $v_\mathrm{g}$ and the net CAI velocities are directed inward. Therefore, CAI velocities in the early phase are positive over a finite range of heliocentric distances only, bounded by two ``stagnation'' points (i.e. heliocentric distances where the CAI velocity is zero).
After some expansion, these stagnation points merge and the velocities are directed inward throughout the disk from that time onward. If, in the self-similar model, the position of the outer stagnation point is taken to be stationary in time, this yields the maximum transport range of solids $R_{\rm max}$ and the corresponding time $t_{\rm max}$\footnote{In this calculation we approximate the temperature with its irradiation contribution, as is appropriate for large heliocentric distances.}:
\begin{equation}
R_{\mathrm{max}}=50\:\mathrm{AU}\:\left(\frac{1\:\mathrm{kg/m^2}}{\rho_s a}\frac{\alpha}{10^{-2}}\frac{M_D R_D^{1/2}}{0.5\:\rm M_{\odot}.\mathrm{AU}^{1/2}}\right)^{2/5}
\end{equation}
\begin{equation}
t_{\rm max}=10^5\:\mathrm{yr}\:\left(\frac{1\:\mathrm{kg/m^2}}{\rho_s a}\frac{M_DR_D^{1/2}}{0.5\:\rm M_{\odot}.\mathrm{AU}^{1/2}}\right)^{2/5}\left(\frac{10^{-2}}{\alpha}\right)^{3/5}\frac{280\:\mathrm{K}}{\rm T_{\mathrm{irr}}(1\:\mathrm{AU})}.
\end{equation}
For fiducial disk parameters, solids can thus be transported outward for tens of AUs within $\sim 10^5$ yr. Because the CAIs must have departed from a sufficiently hot disk region, their transport range may be somewhat smaller than $R_{\mathrm{max}}$ (e.g. $\sim$10 AU in Fig. 2).  
\begin{figure}
\resizebox{\hsize}{!}{\includegraphics{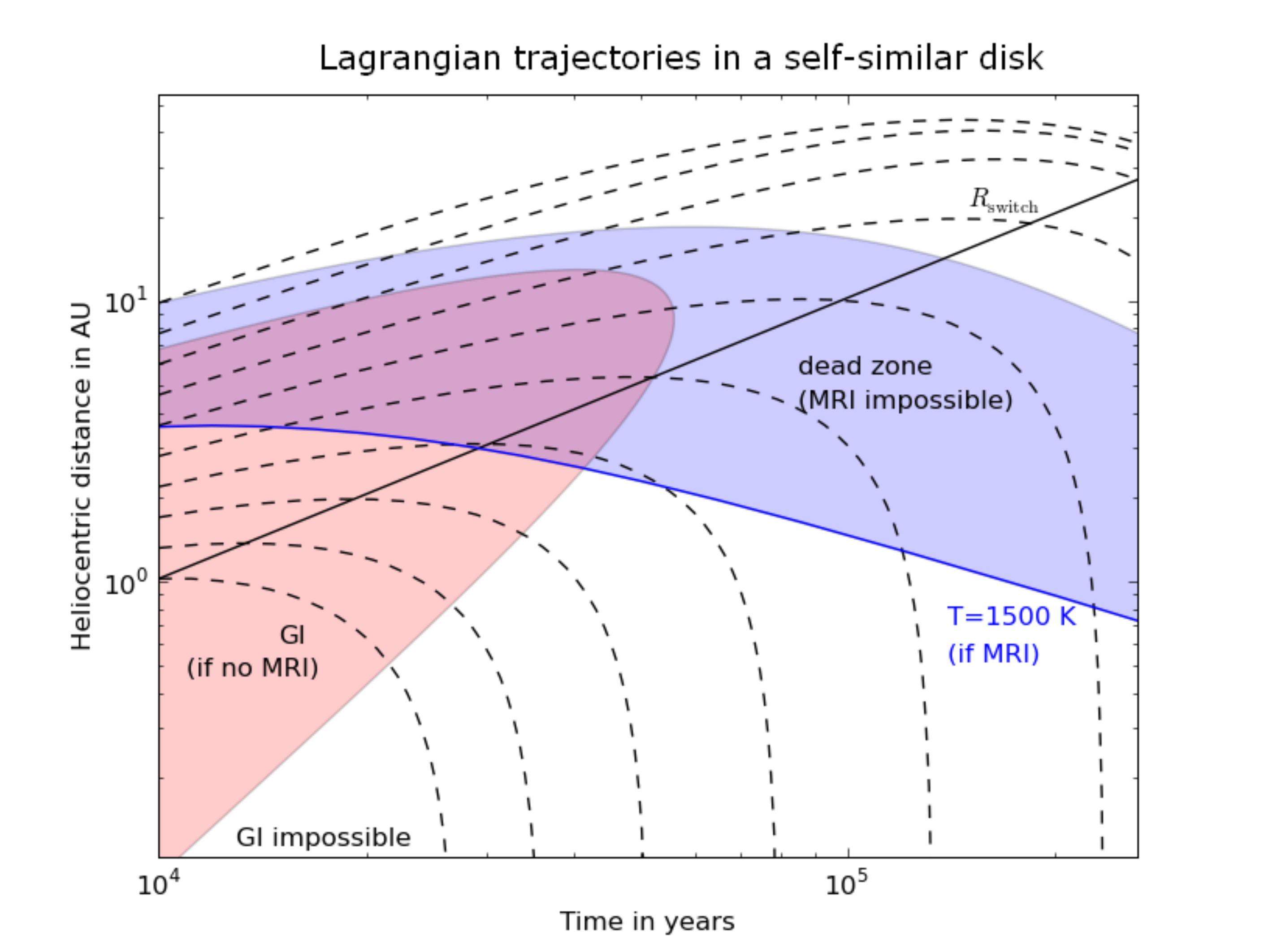}}
\caption{Space-time diagram of the self-similar model. Dashed lines mark the trajectory of CAIs with $\rho_s a=1\:\rm kg/m^2$ and the black continuous line indicates the locus of $R_{\mathrm{switch}}$. The MRI-\textit{inactive} areas are shaded in blue (dead zone), with the 1500 K isotherm (a typical CAI condensation temperature) drawn as a continuous blue line. Regions prone to GI in the \textit{absence} of viscous dissipation are shaded in red. Once the (GI-free) dead zone reaches a significant fraction of the disk mass, the disk model is no longer self-consistent, and disk evolution is expected to slow down, as is the sunward drift of CAIs in the dead zone. The conclusion is that CAIs can be transported to the dead zone before its formation.}
\end{figure}
\section{Preservation of CAIs in the dead zone}
In the self-similar model the return to the Sun, which essentially follows the gas, is completed for $t=4.9\:t_{\rm max}$ \footnote{This result is obtained by numerically integrating the (Lagrangian) equation of motion of the solids at the last stagnation point, i.e. where and when the last positive velocities in the disk disappear.}, that is, largely before chondrite accretion. Although the estimates here are conservative, because they ignore random-walk turbulent motions \citep[e.g.][]{Cuzzietal2003,Ciesla2010}, the problem is especially acute for centimeter-sized CAIs common in CV chondrites (with $\rho_sa$ = 10-100 $\rm kg/m^2$). The purpose of this section is to show that the appearance of a dead zone, which slows down the inward motion of the solids, solves the preservation issue. 

  Indeed, thus far we have not addressed the question of the driver of accretion, and hence the maximum domain of validity of the model. These drivers are thought to result from GI and/or the MRI. The red zone in Fig. 2 shows the maximal domain (calculated ignoring heating by local turbulence) where GI-induced turbulence can occur, assuming a threshold value for the Toomre parameter $Q\equiv\Omega c_s/\pi G \Sigma$ of $Q_c=2$ \citep[see e.g.][]{Kratteretal2008}.
 Although GI may be important in the early phases, MRI-powered turbulence is believed to prevail after the disk has significantly expanded. The blue zone shows where neither temperature (below an activation threshold here fixed at 1500 K) nor cosmic rays (limited to an attenuation depth of $10^3 \:\mathrm{kg/m^2}$) yield the minimum ionization fraction required for the MRI \citep{Gammie1996}. Thus, wherever the blue zone is not overlaid by the red one, $\alpha$ values assumed by the self-similar model ($10^{-3}-10^{-2}$) cannot be self-consistently sustained, and a \textit{dead zone}, with reduced accretion, results within $10^4-10^5$ yr at a few AU (stage 'C' in Fig. 1). \textit{This is before the CAIs have returned to the Sun}. 

  In the evolved stages of the disk (no longer captured by the self-similar model, because its simple $\alpha$ prescription breaks down), the inner edge of the dead zone will move inward to a fraction of an AU, while its outer edge will lie at tens of AU \citep[e.g.][]{Sanoetal2000,IlgnerNelson2006,BaiGoodman2009}. The dead zone would then persist over most of the gas disk's lifetime: \citet{Fromangetal2002} and \citet{Terquem2008} found e.g. dead zones in steady $\alpha-$disks for a wide parameter range, including mass accretion rates at the lower end ($\sim\!\rm 10^{-9}\:M_{\odot}/yr$) of those of observed T Tauri objects \citep{Hartmannetal1998}.

  Accretion is reduced in the dead zone, and the gas density is high enough to also reduce $v_{\mathrm{drag}}$ (stage 'D' in Fig. 1). Using Eq. (\ref{vg}) and (\ref{vdrag}), the CAI drift timescale is
\begin{equation}
t_{\rm drift}\equiv\frac{R}{v_\mathrm{g}+v_{\mathrm{drag}}}\sim\rm 2\:Myr\:R_{\rm AU}^{1/2}\frac{280\:\mathrm{K}}{\mathrm{T}}\frac{10^{-4}}{\mathrm{max}(\alpha,S\!t)}
\label{tdrift}
\end{equation}  
The Stokes number $\mathrm{St}=\pi\rho_s a/2\Sigma$ is related to the gas drag and $\alpha$ to advection. 
 Note that this general order-of-magnitude estimate is independent of the previous history of the disk. $\alpha$ would not exceed $10^{-4}$ in the dead zone \citep{FlemingStone2003} and the surface density would likely be in the range $10^4\!-\!10^5\:\rm kg/m^2$ \citep{Terquem2008}, so that millimeter-sized bodies satisfy $\mathrm{St}\leq 10^{-4}$. This should ensure survival for a few Myr at a few AU.

   After the formation of the dead zone, the CAI production would essentially be confined to the inner active zone. Indeed, an important point is that the temperature of the thermal activation of the MRI is on the order that of CAI condensation. Beyond the inner edge of the dead zone, the temperature will drop significantly below this value because of the drop of the turbulence level. Because diffusion from inside this boundary would be negligible, CAIs formed after the dead zone appearance would not be sent to the chondrite-forming regions in significant amounts.
\section{Discussion}
Using viscous expansion of the disk to transport solids in general \citep{WehrstedtGail2002,Garaud2007,HughesArmitage2010} or specifically CAIs \citep{Ciesla2010}---with the recognition that their \textit{observed} age distribution is narrowed because CAI transport rapidly becomes inefficient---, is not new. However, the input parameters of the runs of \citet{Ciesla2010} always placed the CAI formation region inside $R_{\rm switch}$ and therefore required diffusion to send CAIs to the spreading zone. In Sect. 2 we argued that an initial disk even more compact than envisioned by \citet{Ciesla2010} is conceivable and gave a prime role to advective transport. Its advantage is that the CAI-to-gas ratio would essentially retain its original order of magnitude, which seems required to account for the high mass fraction of CAIs in some chondrites. It would also explain the high proportion of crystalline silicates in comets \citep{Woodenetal2005,HughesArmitage2010}. 
Finally, a small disk size would allow turbulent mixing to achieve a significant initial degree of isotopic (and chemical) homogeneity in it, as is generally observed in meteorites \citep[e.g.][]{CassenSummers1983,YangCiesla2010}.
Regardless of the importance of advection, turbulent diffusion would help outward transport of CAIs and consequently, as far as the latter is concerned, the difference between this (very simplified) study and those of \citet{BockeleeMorvanetal2002,Cuzzietal2003,Ciesla2010,HughesArmitage2010}, is quantitative, not qualitative, in nature. Because this model (or those listed just above) describes the origin and evolution of CAIs disk-wide, one prediction is that the key constraint of the restricted formation interval, hitherto drawn only from CV chondrites, should pertain to CAIs of most if not all chondrite groups. 

   We caution that infall should be taken into account in the future, although we ignored it in this study. This will require dedicated numerical simulations. Conceivably, outward expansion of the bulk of the disk beyond the centrifugal radius and appearance of the dead zone could be linked to a drop of the infall rate: matter supply would then be insufficient to sustain GI and would no longer overwhelm viscous-stress-driven flow past the centrifugal radius \citep[e.g.][]{Zhuetal2010b}. In their protostellar collapse simulations, \citet{Tscharnuteretal2009} also reported outward directed flows near the midplane in the hot inner disk, lasting a few millenia only (possibly because of their high assumed turbulence level), and mentioned the possible relevance for CAI formation.  

  At the end of Sect. 3 we observed that the self-similar model fails to preserve CAIs when extrapolated over 1 Myr. This seems a general property of simple $\alpha$-models. \citet{Ciesla2010} found the discrepancy with observed CAI abundance to be between one and two orders of magnitude. He also noted that lower values of $\alpha$, while slowing down disk evolution, would also reduce CAI production because of the lower temperatures. \citet{Cuzzietal2003} already noted this shortcoming and proposed that the inward drift of meter-sized migrators could maintain a 10-100-fold enhanced abundance of condensable matter above solar in the ``CAI factory'', hereby compensating for their subsequent dilution in the nebula. However, this enhanced metallicity would yield partial pressures of oxygen in excess of those inferred from the $\rm Ti^{3+}/Ti^{4+}$ ratio in CAI pyroxene \citep{Grossman2010} unless the migrators entered the CAI factory with C/O $\gtrsim$ 1 \citep[see Eq. 24 in][]{Grossmanetal2008b}. This value lies far above that of the (chemically primitive) matrices of chondrites though and is consistent with the high end only of the range spanned by comet samples and interplanetary dust particles \citep[e.g.][]{Rietmeijer1998,Sandfordetal2008}.
 Moreover, any enhancement would be limited by the amount of migrators in the outer disk \citep{Ciesla2010}, in particular for relatively compact disks. \citet{Ciesla2010} also argues that preferential settling of CAIs to the midplane 
could enhance their concentration in chondrites. With regard to the CAI drift timescale, it is unclear however when the low values of the surface densities ($\sim\! 10^2\:\mathrm{kg/m^2}$) he implies in his estimate, which are two orders of magnitude below the MMSN at a few AU, would hold, especially in conjunction with low turbulence levels. 

  Our point is that a solution to this problem naturally presents itself when the sources of the turbulence that drives mass accretion are considered. Indeed, as pointed out in Sect. 4, simple $\alpha$-models are not self-consistent throughout the disk extent and the evolution in the context of MRI- and/or GI-driven accretion. The formation of a dead zone is a general, unavoidable feature of accretion disks  \citep{Zhuetal2010b}: for instance, in a steady $\alpha$-disk, the mass accretion rate ($\dot{M}=3\pi\Sigma\nu$) below which $Q$ right outside the inner MRI-active zone, evaluated with the passive irradiation temperature, exceeds $Q_c$ is given by 
\begin{eqnarray}
\dot{M}=2\times 10^{-6}\:\mathrm{M_\odot/yr}\left(\frac{\alpha}{10^{-2}}\right)^{7/8}\left(\frac{2}{Q_c}\right)^{3/4}\left(\frac{\mathrm{T_{MRI}}}{1500\:\mathrm{K}}\right)^{5/8}\nonumber\\\left(\frac{M_\odot}{M_\ast}\right)^{3/16}\left(\frac{\mathrm{T_{\mathrm{irr}}(1\:AU)}}{280\:\mathrm{K}}\right)^{9/8}\left(\frac{0.5\:\mathrm{m^2/kg}}{\kappa}\right)^{1/8},
\end{eqnarray}
which is therefore a lower bound for the appearance of the dead zone in this case. Thus, the preservation mechanism outlined in Sect. 4 is actually fairly independent of the compact disk expansion scenario of Sect. 3. \citet{Ciesla2010} mentioned among alternative possibilities that a slowing of the disk evolution would explain CAI survival ; this study thus provides a way to realize this. That a dead zone blocks diffusion from the inner active zone would also account for the restricted formation interval of CAIs incorporated in chondrites, regardless of the dominating transport mechanism (advection, diffusion...) before the dead zone formation. The presence of a dead zone is also constrained by the Myr spread in age observed for chondrules (whose sizes are similar to that of CAIs) from individual chondrites \citep[e.g.][]{Villeneuveetal2009} by virtue of Eq. (\ref{tdrift}). Finally, although we showed the necessity of the formation of the dead zone, \textit{how} this formation occurs has yet to be ascertained. Because the viscosity drops \textit{outward} around the inner boundary of the dead zone, a net outward flow may be expected there according to the first equality in Eq. (\ref{vg})---perhaps even independently of the compactness of the disk---, before some steady state with uniformly inward accretion sets in. Detailed numerical simulations of dead zone formation are definitely needed.

\begin{acknowledgements}
We thank Dr F. J. Ciesla for his constructive review that significantly improved the quality and the clarity of the manuscript.
\end{acknowledgements}

\bibliographystyle{aa}
\bibliography{16118bibli}
\end{document}